\documentclass[letterpaper, 10 pt, conference]{ieeeconf}  
\usepackage{graphicx}
\usepackage{makecell}
\usepackage{url}
\usepackage{subcaption}
\usepackage{balance}
\usepackage{flushend}
\usepackage{multirow}
\usepackage{layouts}
\usepackage[hidelinks]{hyperref}
\newcommand{\figref}[1]{Fig.~\ref{#1}}
\newcommand{\tabref}[1]{Table~\ref{#1}}
\usepackage{xcolor}
\usepackage{amsfonts}
\usepackage{float}

\IEEEoverridecommandlockouts                              

\title{\LARGE \bf
Leveraging Foundation Models for Calibration-Free c-VEP BCIs 
}

\author{Mohammadreza Behboodi$^{1}$, Eli Kinney-Lang$^{1}$, Ali Etemad$^{2}$, Adam Kirton$^{3}$, Hatem Abou-Zeid$^{4}$
\thanks{*This work was supported by the Alberta Children's Hospital Foundation, the Alberta Graduate Excellence Scholarship, and the NSERC Alliance -  Alberta Innovates Program, all for research funding.}
\thanks{$^{1}$Mohammadreza Behboodi and Eli Kinney-Lang are with the Department of Biomedical Engineering, University of Calgary, 2500 University Drive NW, Calgary, Alberta, Canada
        {\tt\small mohammadreza.behbood@ucalgary.ca},  {\tt\small eli.kinneylang@ucalgary.ca}}%
\thanks{$^{2}$Ali Etemad is with the Department of Electrical and Computer Engineering, Queen’s University, 99 University Ave, Kingston, Ontario, Canada
        {\tt\small ali.etemad@queensu.ca}}%
\thanks{$^{3}$Adam Kirton is with the Department of Pediatric and Clinical Neuroscience, University of Calgary, 2500 University Drive NW, Calgary, Alberta, Canada
        {\tt\small cakirton@ucalgary.ca}}%
\thanks{$^{4}$Hatem Abou-Zeid is with the Department of Electrical and Software Engineering, University of Calgary, 2500 University Drive NW, Calgary, Alberta, Canada
        {\tt\small hatem.abouzeid@ucalgary.ca}}%
\thanks{\textbf{Preprint notice:} This is the authors’ version of the paper accepted for presentation at the \emph{IEEE International Conference on Systems, Man, and Cybernetics (SMC) 2025}, Vienna, Austria. \textcopyright~2025 IEEE. Personal use of this material is permitted. Permission from IEEE must be obtained for all other uses, including reprinting/republishing this material for advertising or promotional purposes, creating new collective works for resale or redistribution to servers or lists, or reuse of any copyrighted component of this work in other works. The final published version will be available on IEEE Xplore.}%
}

\begin{document}

\maketitle
\thispagestyle{empty}
\pagestyle{empty}

\begin{abstract}


Foundation Models (FMs) have surged in popularity over the past five years, with applications spanning fields from computer vision to natural language processing. At the same time, Brain-Computer Interfaces (BCIs) have also gained momentum due to their potential to support individuals with complex disabilities. Among various BCI paradigms, code-modulated Visual Evoked Potentials (c-VEPs) remain relatively understudied, despite offering high information transfer rates and large selection target capacities. However, c-VEP systems require lengthy calibration sessions, significantly limiting their practicality, particularly outside of laboratory settings. In this study, we use a FM for the first time to eliminate the need for lengthy calibration in c-VEP BCI systems. We evaluated two approaches: (1) a truly calibration-free approach requiring no subject-specific data, and (2) a limited calibration approach, where we assessed the benefit of incorporating incremental amounts of calibration data. In both cases, a classification head is trained on data from other subjects. For a new subject, no calibration data is required in the calibration-free setup, making the c-VEP system effectively plug-and-play. The proposed method was tested on two c-VEP datasets. For the calibration-free approach, the average accuracy on the first dataset (n = 17) was $68.8\%\pm17.6\%$, comparable to the full-calibration performance reported in the original study ($66.2\%\pm13.8\%$), which required approximately 11 minutes of calibration. On the second dataset (n = 12), the calibration-free accuracy was $71.8\%\pm20.2\%$, versus $93.7\%\pm5.5\%$ from the original study, which required around 3.5 minutes. A limited-calibration approach using only 20\% of the subject's data (approximately 43 seconds) yielded $92\%\pm5.2\%$ accuracy. These results indicate that our FM–based approach can effectively eliminate or significantly reduce the need for lengthy calibration in c-VEP BCIs.

\end{abstract}

\section{INTRODUCTION}

Brain-Computer Interfaces (BCIs) enable direct communication and control through brain activity, bypassing the need for muscle involvement \cite{Vahid_2023}. Recent interest in BCI technologies has highlighted their potential to improve the quality of life for individuals with complex physical disabilities, such as cerebral palsy \cite{Jadavji_2022}, spinal cord injuries, and locked-in syndrome \cite{Wolpaw_2002}. Among Visual Evoked Potential (VEP) -based BCIs, code-modulated VEP (c-VEP) stands out due to its ability to present numerous visual stimuli (64 in \cite{Liu_2018}) and achieve high classification accuracy \cite{Bin_2009}. In c-VEP BCIs, users focus on visual stimuli that flash in binary sequences with specific patterns. These patterns evoke distinct electrical responses in the brain, known as VEPs, which are recorded using EEG. The system analyzes these VEPs to determine which stimulus the user is attending to, enabling tasks such as robot control \cite{Behboodi_2020, Waytowich_2015}.

C-VEP BCIs have the potential to empower children with limited motor abilities by interpreting their brain responses to visual cues, thereby enabling interaction with their environment and supporting overall development \cite{Orlandi_2021,Jadavji_2022}. Prior work has shown that classification accuracy in c-VEP BCIs heavily depends on the availability of sufficient training EEG data for each individual \cite{Behboodi_2024, Martinez, Wei_2018, Martinez_2021}. However, obtaining such data is especially challenging in pediatric populations. Children with complex needs often face barriers to using BCI systems \cite{KinneyLang_2023}, in part due to long and complex calibration phases. This may help explain the lack of c-VEP BCI studies involving children, despite their potential for empowering these users through the flexible and rapid c-VEP control paradigm. Due to these lengthy calibration procedures, c-VEP BCIs remain largely inaccessible for this population.

Differences in brain responses across users and sessions further complicate generalization of c-VEP models, often requiring subject-specific calibration before use \cite{Martinez_2021, Guetschel_2024}. Coupled with the visual fatigue caused by the rapidly flashing stimuli in c-VEP systems, long calibration sessions can negatively impact user comfort and system performance in real-world applications \cite{Ajami_2018, Azadi_2023}. Reducing or eliminating calibration time has the potential to decrease visual and mental fatigue, thereby improving both comfort and performance \cite{Chai_2020}.

To address these challenges, we propose the use of a Foundation Model (FM) pre-trained on large-scale adult EEG datasets from BCI paradigms other than c-VEP (see \tabref{tab:Tab1}). This FM is used as a frozen encoder, while a lightweight task head is trained on c-VEP data. We evaluate three training strategies for the task head: (1) a fully calibration-free approach, (2) limited calibration using a small portion of a new subject's data, and (3) within-subject calibration from scratch. The evaluation was conducted on two c-VEP datasets: Fast-Stim \cite{Wittevrongel_2017} and Group-Mod \cite{Wei_2018}. For the Fast-Stim dataset, excluding one outlier subject, the calibration-free approach achieved higher accuracy than the full-calibration method reported in the original study (71.8\% vs. 67.4\%). For the Group-Mod dataset, the calibration-free accuracy was lower than that of the original study (71.8\% vs. 93.7\%), which required 214 seconds of calibration data. However, using the limited calibration approach, we achieved comparable accuracies of 89.5\% and 92\% with only 22 and 43 seconds of calibration data, respectively.

Our contributions in this paper include:
\begin{itemize}
    \item We demonstrate, for the first time, that a FM can enable a calibration-free c-VEP BCI while maintaining competitive accuracy of greater than 70\% in comparison to standard calibration-based methods.

    \item In cases where calibration-free performance falls short of full calibration, we show that fine-tuning with a small amount of subject-specific data (less than one-fifth of the data used in full calibration) effectively bridges the performance gap.

    \item We show that incorporating data from other subjects when training the task head consistently improves classification accuracy for an unseen subject, compared to training solely on that subject’s data.

    \item The proposed approaches are compatible with various types of stimulus sequences and calibration strategies and do not depend on any specific sequence design.
\end{itemize}

\section{Related Work}

\subsection{Background on c-VEP BCI}
In c-VEP BCI systems, two primary calibration strategies are commonly used. The first, known as the Circular Shift approach, assigns a pseudorandom sequence to a reference visual target. Circularly shifted versions of this sequence are then allocated to the remaining targets. During calibration, the subject repeatedly focuses on the reference target while EEG signals are recorded. Responses for other targets are generated by applying circular shifts to the reference response \cite{Behboodi_2020}. This approach offers the advantage of reduced calibration time \cite{Martinez_2021}. The second strategy, often referred to as the Ensemble method, uses stimulus sequences that may be either independent or circularly shifted versions. In this approach, subjects attend to each target individually to directly record stimulus-specific EEG responses \cite{Wittevrongel_2017}. The Ensemble method typically yields higher classification accuracy, as each target is associated with a uniquely measured brain response \cite{Martinez_2021}.

\subsection{Calibration in c-VEP BCIs}
Several recent studies have attempted to reduce or eliminate calibration time in c-VEP BCIs, with varying levels of success. Thielen et al. \cite{Thielen_2021} introduced a reconvolution-based encoding model that eliminated calibration by modeling the brain’s response to Gold code stimuli using generalizable linear filters. Zheng et al. \cite{Zheng_2024} proposed Narrow-Band Random Sequences (NBRS) combined with Filter-Bank Canonical Correlation Analysis (FBCCA), enabling decoding based solely on stimulus design. 

Further work by Thielen et al. \cite{Thielen_2024} introduced Unsupervised Mean-difference Maximization (UMM) and cumulative Canonical Correlation Analysis (CCA), achieving competitive performance without labeled data. Castillos et al. \cite{Kalou_2023} used a CNN-based bitwise decoding method within a burst c-VEP paradigm, reducing calibration to just a few seconds. Miao et al. \cite{Yining_2024} developed a high-performance White-Noise (WN) c-VEP BCI with a brief single-target calibration, leveraging linear models and transfer learning to achieve performance comparable to traditional SSVEP systems. Huang et al. \cite{HUANG_2020} proposed a transfer learning framework that uses data from selected source subjects to build a model for a target subject, based on a similarity criterion.

While each of these studies presents innovative strategies to shorten or eliminate calibration, most rely on custom stimulus designs or still require some subject-specific data. In contrast, our study introduces the use of a FM that generalizes across stimulus types and subjects, offering a robust, calibration-free c-VEP solution.

\subsection{Foundation Models (FMs) in BCIs}
In the context of BCIs, FMs are large neural networks pre-trained on vast, unlabeled EEG datasets. These models can be fine-tuned for specific downstream classification tasks using minimal labeled data \cite{Guetschel_2024, Abibullaev_2023}. Kim et al. \cite{Kim_2024} developed EEG-GPT, a model pre-trained on the Temple University Hospital (TUH) EEG dataset, which achieved high accuracy on downstream tasks using only 2\% of labeled data for fine-tuning. Other studies have introduced similar models, including BENDR \cite{Kostos_2021}, BIOT \cite{Chaoqi_2023}, NeuroGPT \cite{Cui_2024}, and LaBraM \cite{jiang_2024}. These models leveraged large-scale clinical and BCI EEG datasets spanning various domains, including sleep staging, motor imagery, and emotion recognition. Compared to training deep models from scratch, FMs have demonstrated significant performance gains, particularly in low-data scenarios (e.g., over 4\% improvement reported in \cite{Cui_2024, jiang_2024}), highlighting their strong generalization capabilities. Most recently, Wang et al. introduced EEGPT \cite{Wang_2024}, which established a new state-of-the-art by being pre-trained on a large, multi-task EEG dataset (see \tabref{tab:Tab1}).

While these approaches have improved BCI performance across many paradigms, none have specifically addressed the c-VEP decoding task. Our study is the first to apply a FM to eliminate or minimize calibration in c-VEP BCIs, thereby addressing this gap.

\section{Methodology}

\subsection{Problem Statement}
Let $\mathbf{X} \in \mathbb{R}^{C \times T}$ represent the EEG response of a subject while attending to a c-VEP visual stimulus, where $C$ is the number of EEG channels and $T$ is the number of time samples. A c-VEP BCI aims to learn a function $f(\mathbf{X}) = Y$, where $Y$ denotes the index of the target visual stimulus detected by the system. Traditional c-VEP BCIs require long calibration sessions to accurately learn the function $f$, which may lead to fatigue, boredom, and disengagement for users. This study aims to eliminate or reduce the calibration time by leveraging a FM to enable a calibration-free or calibration-minimized c-VEP BCI system.

\subsection{Our Approach}
We utilize the state-of-the-art EEGPT FM \cite{Wang_2024} as a frozen encoder and add a new task head composed of two linear layers for classification of EEG responses $\mathbf{X}$ during c-VEP BCI operation. The overall framework is depicted in \figref{fig:Fig1}.
The task head is first trained using a set of calibration data from other subjects. For a new (unseen) subject, we evaluate two approaches:
\begin{itemize}
    \item \textbf{Calibration-free} approach, where the trained task head is directly applied to the new subject without requiring any subject-specific data;
    \item \textbf{Limited calibration} approach, where the task head is fine-tuned using a small subset of the new subject’s calibration data.
\end{itemize}

To evaluate these methods, we use two publicly available c-VEP datasets that follow different paradigms: the Fast-Stim dataset \cite{Wittevrongel_2017} based on the Ensemble approach, and the Group-Mod dataset \cite{Wei_2018} based on the Circular Shift approach.
 
\subsection{Datasets}

\subsubsection{Fast-Stim Dataset}
This publicly available dataset, described in \cite{Wittevrongel_2017}, contains c-VEP recordings from 17 participants with normal or corrected-to-normal vision (14 females, 13 right-handed, mean age $22.35 \pm 2.9$ years, ranging from 18 to 30). Visual stimuli were presented using a 63-bit m-sequence and its circularly shifted versions, with a total of 32 visual targets shown on a 60 Hz screen. EEG was recorded from 32 channels (Fz, C1, Cz, C2, TP7, CP5, CP3, CP1, CPz, CP2, CP4, CP6, TP8, P7, P5, P3, P1, Pz, P2, P4, P6, P8, PO3, POz, PO4, PO7, O1, Oz, O2, PO8), using a SynampsRT system (Compumedics Neuroscan, Austria), sampled at 2000 Hz. The ground electrode was placed at AFz and the reference at FCz. In each session, participants focused on each visual stimulus five times in randomized order, with each target flashed five times consecutively. This resulted in 800 trials per subject, each with a duration of 1.05 seconds.

\subsubsection{Group-Mod Dataset}
In the Group-Mod dataset (described in \cite{Wei_2018}), EEG signals were recorded from 12 participants (aged 21–26 years) using 9 channels: P3, Pz, P4, PO7, POz, PO8, O1, Oz, and O2. Recordings were performed with a Synamps2 system (Neuroscan Inc.) at a 1000 Hz sampling rate. The dataset includes 16 visual targets, each associated with a Golay 1 code or one of its circular shifts. Stimuli were presented on a 60 Hz monitor. During calibration, each participant was instructed to focus on a reference visual target for 1.06 seconds (sequence duration), repeated across 200 trials. In the testing session, participants viewed each of the 16 visual targets for five repetitions, yielding 80 EEG trials per subject used for evaluation in this study.

\subsection{Representation Extraction via Frozen Encoder}

\begin{figure*}[htbp]
\centering
\makebox{
\includegraphics[width=0.95\textwidth]{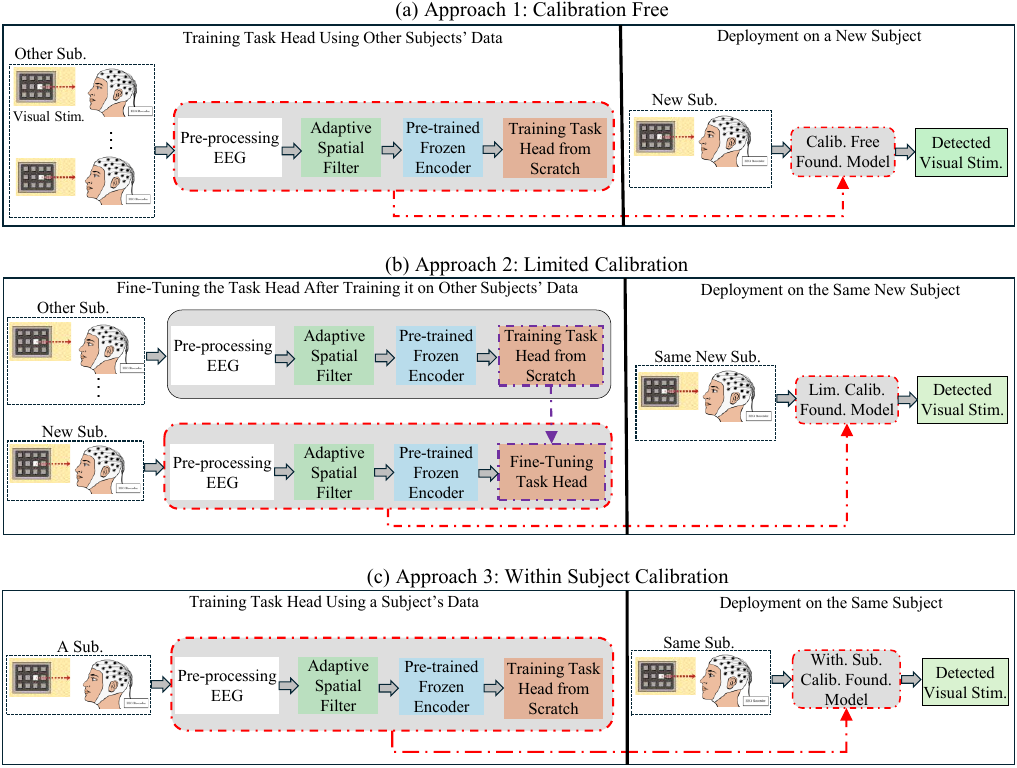}  
}
\caption{The signal processing pipeline for applying the EEGPT foundation model to c-VEP datasets under three scenarios: (a) Calibration-Free, (b) Limited Calibration, and (c) Within-Subject Calibration. For each case, the left panel illustrates task head training, while the right panel shows the deployment stage, which was evaluated offline to simulate real-time classification.}
\label{fig:Fig1}
\end{figure*}

As illustrated in \figref{fig:Fig1}, raw multichannel EEG signals were preprocessed (details in Section IV-A). The preprocessed EEG data are represented as a 3D tensor $\mathbf{X}' \in \mathbb{R}^{B \times C \times T'}$, where $B$ is the batch size (number of trials), $C$ is the number of channels, and $T'$ is the number of time samples per trial (fixed at 1024). An adaptive spatial filter aligns the dataset’s EEG channels to the EEGPT encoder’s input configuration, outputting tensor $\mathbf{X}''$ of the same dimensions. This filtered tensor $\mathbf{X}''$ is passed into the frozen pre-trained EEGPT encoder to extract a deep feature representation $\mathbf{Z} \in \mathbb{R}^{B \times F_1 \times F_2 \times D}$, where $F_1$, $F_2$, and $D$ denote the feature map dimensions and embedding size. In our setup, the encoder output has dimensions $\mathbb{R}^{B \times 16 \times 4 \times 512}$.

\subsection{Task Head}

We adopt the same task head structure used in the original EEGPT paper \cite{Wang_2024}. The task head, appended to the encoder, consists of two linear layers separated by flattening operations. The 4D tensor $\mathbf{Z}$ is first flattened to  $\mathbf{Z}' \in \mathbb{R}^{B \times 16 \times 2048}$. A fully connected layer with 16 neurons maps it to $\mathbf{Z}'' \in \mathbb{R}^{B \times 16 \times 16}$. Another flattening operation produces a 2D feature matrix $\mathbf{H} \in \mathbb{R}^{B \times 256}$, which is fed into a final fully connected layer that outputs classification logits. The number of output neurons corresponds to the number of visual targets: 32 for the Fast-Stim dataset and 16 for the Group-Mod dataset.

\begin{table}[htbp]
\caption{Datasets used for pre-training the EEGPT foundation model, based on \cite{Wang_2024}}
\label{tab:Tab1}
\centering
\begin{tabular}{cccc}
\hline
Dataset & Paradigms & Subject & Targets\\
\hline\hline
Physio MI & MI \& ME & 109 & 5\\
HGD & MI & 14 & 4\\
TSU & SSVEP & 35 & 40\\
SEED & EMO & 15 & 3\\
M3CV & MULTI & 106 & -\\
\hline
\end{tabular}
\end{table}

\subsection{Evaluation Paradigms}
We propose and evaluate three approaches for utilizing the EEGPT FM to remove or reduce c-VEP calibration requirements. Each approach was implemented on both Fast-Stim and Group-Mod datasets to ensure compatibility with both calibration paradigms: Ensemble and Circular Shift.

\subsubsection{Calibration-Free c-VEP}
As shown in \figref{fig:Fig1}(a), in this approach, the task head is trained using pooled data from all subjects except one. The trained task head is then evaluated on the held-out subject. This procedure is repeated for all subjects using a Leave-One-Subject-Out (LOSO) strategy. The model is not exposed to any data from the test subject during training or validation, representing a true plug-and-play scenario.

\subsubsection{Limited Calibration c-VEP}
As shown in \figref{fig:Fig1}(b), the task head is first trained on pooled data from all subjects excluding the target subject. It is then fine-tuned using a limited amount of calibration data from the excluded subject. We evaluated model performance using varying calibration sizes. This process was applied to both datasets.

\subsubsection{Within-Subject Calibration c-VEP}
Shown in \figref{fig:Fig1}(c), this approach trains the task head from scratch using only data from the target subject, with no pre-training on other subjects. As with the limited calibration setup, we evaluated model performance using increasing portions of the subject's data.

\section{Experimental Setup}
\subsection{Pre-processing Data for Training Task Head}
For the Fast-Stim dataset, we used the preprocessed version provided by the authors \cite{Wittevrongel_2017}, which had been filtered using a 4th-order Butterworth filter between 4–31 Hz and downsampled to 100 Hz. To assess the impact of trial averaging, we averaged every five EEG trials to generate a single composite trial and evaluated classification accuracy using both individual and averaged trials.

For the Group-Mod dataset, we extracted EEG trials from the raw signals and applied a bandpass filter with cutoff frequencies of 2 and 30 Hz. Trials related to both calibration and testing sessions were segmented from the continuous EEG data. During calibration, responses to the reference visual stimulus were recorded. EEG responses for the remaining stimuli were generated via circular shifting. The resulting trials for all visual stimuli were then pooled and treated as the calibration data for each subject.

\subsection{Evaluation}
All experiments were conducted on a machine equipped with an NVIDIA GeForce RTX 4080 GPU (16 GB RAM). All models were implemented in Python version 3.12.

\subsubsection{Calibration-Free Approach Validation}

For the calibration-free approach, 90\% of the available data (excluding the target subject) was used for training the task head of the FM, while 10\% was reserved for validation. Training was performed for 200 epochs, and the model with the lowest validation loss was selected. No data from the target subject was used during training or validation. The selected model was then tested on the target subject’s data, representing a true subject-independent evaluation.

\subsubsection{Limited Calibration Approach Validation}

In the limited calibration setting, the best model selected from the calibration-free training (for each subject) was used as the starting point. This model was fine-tuned using various proportions of the target subject’s calibration data.

For the Fast-Stim dataset, fine-tuning was performed using 10\%, 20\%, 40\%, 60\%, and 70\% of the target subject's calibration data, with an additional 10\% reserved for validation in each case. Training was conducted over 400 epochs, and the model with the lowest validation loss was selected. Classification accuracy was then evaluated using a 20\% hold-out test set. For comparison, \figref{fig:Fig2} also presents the calibration-free results, where only 20\% of the target subject’s data was used for calculating accuracy.

For the Group-Mod dataset, a similar approach was used, with 10\%, 20\%, 40\%, 60\%, and 80\% of the calibration data used for fine-tuning. As before, 10\% of the calibration data was held out for validation, and the entire test session data was used for evaluation.

\subsubsection{Within-Subject Approach Validation}

The within-subject calibration approach followed the same procedure and hyperparameters as the limited calibration setup. However, in this case, the task head was trained from scratch using only the target subject’s data. Classification accuracy was evaluated similarly for comparison.

\section{Results \& Discussion}

\subsection{Calibration Free Approach Results}
The results of the calibration-free approach on the Fast-Stim dataset are shown in \tabref{tab:Tab2}. As seen in the table, subject 11 exhibited significantly lower accuracy compared to other participants and was therefore considered an outlier. Consequently, results are reported both including and excluding subject 11.

Excluding subject 11, the mean accuracy for the calibration-free approach using 1-second trials was $71.8\%\pm13.3\%$. When the EEG trial duration was increased to 5 seconds, the accuracy improved to $94.3\%\pm6.9\%$. These results are comparable to those reported in the original study, which achieved $67.4\%\pm13.3\%$ and $98.7\%\pm2.1\%$ for 1-second and 5-second trials, respectively. The original study required 672 seconds of calibration and used a spatio-temporal beamformer for classification.

For the Group-Mod dataset, results of the calibration-free approach are presented in \tabref{tab:Tab3}. The mean classification accuracy for 1-second trials using our method was $71.8\%\pm20.17\%$. In contrast, the original study achieved $93.7\%\pm5.5\%$, but required 214 seconds of calibration per subject.

This discrepancy can be attributed to the way calibration data was recorded. In the Group-Mod dataset, only EEG responses to a reference visual stimulus were recorded during calibration. Responses to the other stimuli were synthesized using circular shifts. As a result, the training data available to the FM consisted of actual responses only to the reference stimulus, while responses to other targets were estimated—possibly limiting the model’s ability to generalize.

Nevertheless, as discussed in the next subsection, the performance of the FM improves significantly with the inclusion of a small amount of real calibration data. In the original study of the Group-Mod dataset, classification was performed using CCA-based spatial filtering and a template-matching approach.

\begin{table}[H]
\caption{Classification accuracies and calibration time for each subject using the Calibration-Free approach, compared to the results reported in the original Fast-Stim dataset study.}
\label{tab:Tab2}
\centering
\footnotesize 
\setlength{\tabcolsep}{2pt} 
\begin{tabular}{ccc|cc}
\hline
\textbf{Sub.} & \multicolumn{2}{c|}{\textbf{Calibration Free}} & \multicolumn{2}{c}{\textbf{Original Study}} \\
\hline
& \makecell{\textbf{1 sec. trial} \\ \textbf{Acc. (\%)}} & \makecell{\textbf{5 sec. trial} \\ \textbf{Acc. (\%)}} & \makecell{\textbf{1 sec. trial} \\ \textbf{Acc. (\%)}} & \makecell{\textbf{5 sec. trial} \\ \textbf{Acc. (\%)}} \\
\hline\hline
1 & 75 & 98.1 & 63.1 & 99.4\\
2 & 71.3 & 96.3 & 76.3 & 100\\
3 & 60.8 & 93.8 & 62.5 & 100\\
4 & 78 & 98.8 & 69.4 & 99.4\\
5 & 84.4 & 96.9 & 76.3 & 100\\
6 & 52.6 & 80.6 & 68.1 & 97.5\\
7 & 67.9 & 98.1 & 42.5 & 92.5\\
8 & 45.6 & 81.3 & 56.9 & 98.1\\
9 & 63 & 88.1 & 95.6 & 100\\
10 & 68.5 & 95 & 49.4 & 95\\
11 & 20.8 & 41.9 & 46.3 & 93.1\\
12 & 75 & 93 & 66.3 & 100\\
13 & 93.1 & 100 & 82.5 & 100\\
14 & 84.6 & 99.4 & 59.4 & 100\\
15 & 92.9 & 100 & 66.9 & 99.4\\
16 & 57.5 & 90 & 56.9 & 98.1\\
17 & 78.8 & 99.4 & 86.3 & 100\\
\hline
\makecell{\textbf{Avg.$\pm$Std.} \\ \textbf{Acc. (\%)}} & $68.8 \pm 17.6$ & $91.2 \pm 13.7$ & $66.2 \pm 13.8$ & $98.4 \pm 2.4$ \\
\hline
\makecell{\textbf{Avg.$\pm$Std.} \\ \textbf{(Excl. Sub. 11)} \\ \textbf{Acc. (\%)}} & $71.8 \pm 13.3$ & $94.3 \pm 6.9$ & $67.4 \pm 13.3$ & $98.7 \pm 2.1$\\ 
\hline
\makecell{\textbf{Calib. Time} \\ \textbf{Per Sub. (Sec.)}} & 0 & 0 & 672& 672\\
\hline
\end{tabular}
\end{table}

\begin{table}[H]
\caption{Classification accuracies and calibration time for each subject using the Calibration-Free approach, compared to the results reported in the original Group-Mod dataset study.}
\label{tab:Tab3}
\centering
\footnotesize 
\setlength{\tabcolsep}{2pt} 
\begin{tabular}{ccccc}
\hline
\textbf{Sub.} & \multicolumn{3}{c}{\textbf{Proposed Approaches}} & \multirow{2}{*}{\makecell{\textbf{Original}\\ \textbf{Study}}}\\
\cline{2-4}
& \textbf{Calib. Free} & \makecell{\textbf{Lim. Calib.} \\ \textbf{(10\%)}} & \makecell{\textbf{Lim. Calib.} \\ \textbf{(20\%)}}\\

\hline\hline
1 & 90 & 95 & 96.2 & 88.8\\
2 & 50 & 76.2 & 83.7 & 97.5 \\
3 & 73.8 & 91.3 & 96.2 & 98.8\\
4 & 93.8 & 98.8 & 98 & 93.8\\
5 & 80 & 93.8 & 91.3 & 92.5\\
6 & 66.3 & 86.3 & 88 & 98.8\\
7 & 31.3 & 80 & 90 & 82.5\\
8 & 70 & 93 & 93.8 & 95\\
9 & 98.8 & 98.8 & 98.8 & 100\\
10 & 42.5 & 76.2 & 83.7 & 90\\
11 & 78.8 & 88.7 & 88 & 98.8\\
12 & 86.3 & 96.2 & 98 & 87.5\\
\hline
\makecell{\textbf{Avg.$\pm$Std.} \\ \textbf{Acc. (\%)}} & $71.8 \pm 20.17$ & $89.5 \pm 7.8$ & $92 \pm 5.2$ & $93.7\pm5.5$\\
\hline
\makecell{\textbf{Calib. Time} \\ \textbf{Per Sub. (Sec.)}} & 0 & 22 & 43 & 214 \\
\hline
\end{tabular}
\end{table}

\captionsetup[figure]{skip=4pt}
\begin{figure*}[htbp]
    \centering
    \makebox{
        \begin{minipage}{0.95\textwidth}
            \centering
            \begin{subfigure}{\textwidth}
                \centering
                 \caption{Accuracy vs Calibration Data Percentage for Fast-Stim Dataset}
                 \includegraphics[width=\textwidth]{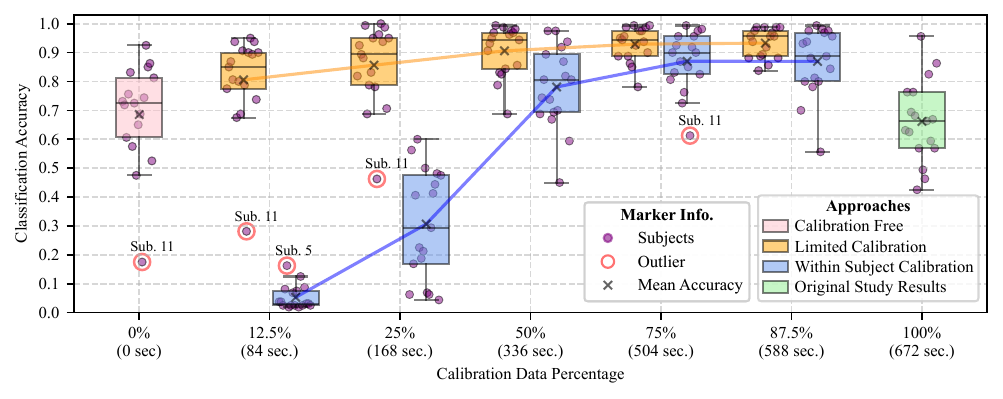}
                 \label{fig:subfig1}
            \end{subfigure}
            
            \vspace{-1.3em}  
            
            \begin{subfigure}{\textwidth}
                \centering
                \caption{Accuracy vs Calibration Data Percentage for Group-Mod Dataset}
                 \includegraphics[width=\textwidth]{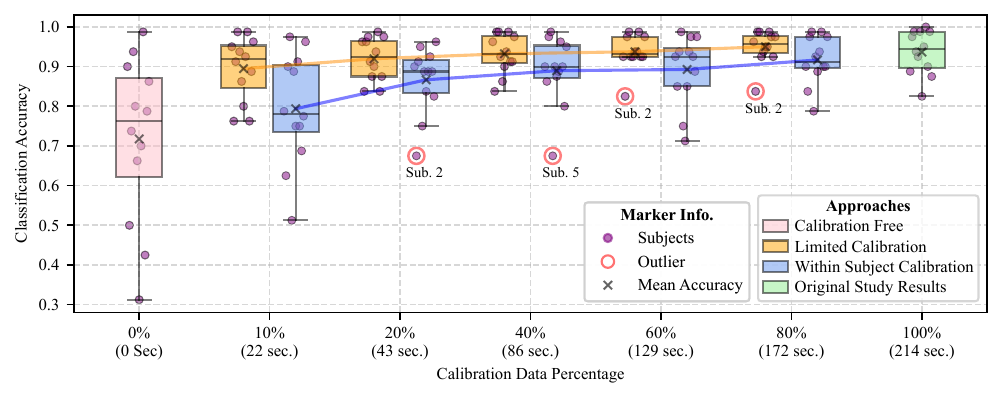}
                 \label{fig:subfig2}
            \end{subfigure}
        \end{minipage}
    }
    \vspace{-1.2em}  
    \caption{Box plots comparing the results of the Calibration-Free, Limited Calibration, and Within-Subject Calibration approaches for (a) the Fast-Stim dataset and (b) the Group-Mod dataset. For the Limited and Within-Subject Calibration approaches, classification accuracy is shown across varying percentages of calibration data used to train the task head. For reference, the results from the original study for each dataset are also included.} 
    \label{fig:Fig2}
\end{figure*}

\subsection{Limited Calibration \& Within-Subject Calibration Approaches Results}
\figref{fig:Fig2} presents box plots comparing the Limited Calibration and Within-Subject Calibration approaches. In the Limited Calibration approach, the task head of the pre-trained FM was fine-tuned using various portions of the target subject’s calibration data. In contrast, the Within-Subject Calibration approach involved training the task head from scratch using only the target subject’s data. The results of both original studies are also shown for reference. 

For the Fast-Stim dataset (\figref{fig:Fig2}(a)), using only 12.5\% (84 seconds) of the target subject’s calibration data, the mean accuracy in the Limited Calibration approach increased from $68.5\%\pm17.4\%$ (Calibration-Free) to $80.6\%\pm15.5\%$, outperforming the original study ($66.2\%\pm13.8\%$). With 75\% of the data (504 seconds), the mean accuracy reached $93.3\%\pm4.9\%$. Across all percentages, Limited Calibration consistently outperformed Within-Subject Calibration, especially in the 0–50\% range, with differences exceeding 12.5\%. This performance gap narrowed with higher calibration ratios.

For the Group-Mod dataset (\figref{fig:Fig2}(b)), using only 10\% of calibration data (22 seconds), the mean accuracy improved from $71.8\%\pm20.2\%$ to $89.5\%\pm7.8\%$. Increasing the calibration portion to 80\% (172 seconds) resulted in $95.0\%\pm4.1\%$, exceeding the $93.7\%\pm5.5\%$ reported in the original study, which used 100\% of the data (214 seconds). In all cases, the Limited Calibration approach achieved higher accuracy than Within-Subject Calibration, with a notable margin at 10\% (89.5\% vs. 79.4\%). This gap shrank to less than 5.5\% at 20\% of calibration data and above.

\subsection{Comparison with other related studies}
This study introduces the first use of a FM for minimizing or eliminating the calibration time of c-VEP BCIs. The FM was pre-trained using large-scale EEG data from non–c-VEP paradigms (see \tabref{tab:Tab1}), yet performed effectively when fine-tuned using only a small c-VEP dataset.

Importantly, the two evaluated datasets—Fast-Stim and Group-Mod—differ in both stimulus sequence design (m-sequence vs. Golay 1) and calibration strategy (Ensemble vs. Circular Shift). Our model generalized well across both, demonstrating its robustness to different experimental setups and sequence types.

Compared to prior c-VEP studies focused on calibration reduction (see Section II-B), our approach is the first to demonstrate a true calibration-free solution without depending on specific stimulus sequences or calibration paradigms. For example, Castillos et al. \cite{Kalou_2023} focused only on reducing calibration time using a custom burst code. Huang et al. \cite{HUANG_2020} proposed transfer learning, but still required some target subject data to identify suitable source subjects. Zheng et al. \cite{Zheng_2024} proposed a calibration-free method based on narrow-band stimulus sequences, which limits its general applicability. Finally, reconvolution-based approaches by Thielen et al. \cite{Thielen_2021, Thielen_2024} were designed specifically for Gold Code sequences and have not been generalized to other encoding schemes.

\subsection{Study Limitations \& Next steps}
One of the high-level motivations behind this study was to make c-VEP BCIs more suitable for pediatric users by eliminating the need for lengthy calibration. However, the datasets used for evaluation consisted solely of adult participants. To the best of our knowledge, no publicly available c-VEP datasets currently exist for pediatric populations. Nevertheless, we believe that the outcomes of this pilot study offer valuable insights that may generalize to pediatric users. As part of future work, we aim to collect a dedicated pediatric c-VEP dataset and directly evaluate the proposed approaches in that context.

The success of our Calibration-Free and Limited Calibration approaches depends on access to a dataset with sufficient numbers of subjects recorded using the same experimental setup. Another limitation is the computational complexity and memory requirements of the EEGPT FM, which may hinder deployment on low-resource hardware. Future work will explore using alternative FMs and fine-tuning strategies to further improve generalizability and reduce model size.

\section{Conclusion}
In this study, we applied a pre-trained FM to c-VEP BCIs for the first time to remove or minimize the need for lengthy calibration. Results on the Fast-Stim dataset showed that our calibration-free approach achieved accuracy levels comparable to those reported in the original study, which required extensive calibration. For the Group-Mod dataset, we demonstrated that while calibration-free performance was reasonable, using just 20\% of the subject’s calibration data enabled our model to match the accuracy of the original full-calibration approach.

These findings support the feasibility of developing calibration-free c-VEP BCIs by leveraging FMs trained on large-scale public EEG datasets and adapting them with only a small amount of task-specific c-VEP data. Our approach represents an important step toward making c-VEP BCIs more practical and accessible, particularly for end users such as children or adults with disabilities and complex neurological conditions.



\bibliographystyle{IEEEtran}
\bibliography{references}

\end{document}